\documentclass[allclo]{FBSart}
\usepackage{epsfig}

\def\lb {\left[ }
\def\rb {\right] }

\def\ra {\rangle }
\def\la {\langle }

\def\rar {\rightarrow}

\def\beq{\begin{equation}}
\def\eeq{\end{equation}}
\def\bea{\begin{eqnarray}}
\def\eea{\end{eqnarray}}

\def\cL {{\cal{L}}}

\def\cO{{\cal{O}}}

\def\D {\Delta}

\def\m{\mu}

\def\p {\pi}

\def\s{\sigma}

\def\ub {\bar u}

\def\sp {\!+\!}
\def\sm {\!-\!}

\def\br {\mbox{\boldmath $r$}}

\def\bnb {\mbox{\boldmath $\nabla$}}

\title{Scalar Form Factors\\
\hspace*{48mm} and Nuclear Interactions}
\author{\underline{M. R. Robilotta} and G. R. S. Zarnauskas
}
\institute{Instituto de F\'{\i}sica, Universidade de S\~ao Paulo,
S\~ao Paulo, SP, Brazil}

\runningauthor{M.\,R.\,Robilotta and G. R. S. Zarnauskas}
\runningtitle{Scalar Form Factors and Nuclear Interactions}
\begin{document}

\maketitle
\begin{abstract}
The scalar-isoscalar term in the two-pion exchange $NN$ potential
is abnormally large and does not respect the hierarchy of effects 
predicted by chiral perturbation theory.
We argue that this anomaly is associated with non-perturbative effects,
which are also present in the $\pi N$ scalar form factor.
\end{abstract}

\section{Nuclear Potentials}

In the last fifteen years, chiral perturbation theory (ChPT) has been 
systematically employed in the study of nuclear interactions.
In ChPT, amplitudes are expanded in power series of a typical variable $q$, 
representing either pion four-momenta or nucleon three-momenta, 
such that $q<<1$ GeV. 
In the case of nuclear interactions, ChPT predicts a hierarchy 
in which potentials have structures with different orders of
magnitude. 

The leading contribution to the $NN$ potential begins at $\cO(q^0)$ and is 
due to the $OPEP$\cite{Bira}.
The two-pion exchange contribution $(TPEP)$ begins at $O(q^2)$ and has already
been expanded up to $O(q^4)$, by means of both heavy baryon\cite{HB}
and covariant\cite{RHR} ChPT.
The leading contribution to the three-nucleon force, associated with two-pion
exchange, begins at $\cO(q^3)$ and $\cO(q^4)$ corrections are presently being 
evaluated\cite{NNN}. 
Nowadays, about 20 nuclear-force components are known and the overall
picture can be assessed.

An outstanding problem in the hierarchy predicted by ChPT 
concerns the relative sizes of the isospin independent $(V_C^+)$
and dependent $(V_C^-)$ central components of the $TPEP$\cite{HRR}, 
displayed in Fig.1. 
According to ChPT, the former begins at $\cO(q^3)$ and the latter at $\cO(q^2)$.
On the other hand, the figure shows that the chiral 
hierarchy is defied, since 
$V_C^+ \sim 10 \, |V_C^-|$.
These profile functions were scrutinized in Ref.\cite{HRR} and the function 
$V_C^+$ was found to be heavily dominated by a term of the form
\beq
V_C^+ \sim -\, (4/f_\p^2)\;\lb (c_3 - 2c_1) - c_3 \; \bnb^2/2 \rb \;
\tilde{\s}_{N_N}\;,
\label{e1}
\eeq
where the $c_i$ are LECs and $\tilde{\s}_{N_N}$ is the leading contribution 
of the pion cloud to the nucleon scalar form factor.
This close relationship between $\tilde{\s}_{N_N}$ and $V_C^+$ indicates 
that the study of the former can shed light into the properties of the 
latter.
\begin{figure}[hbt]
\epsfig{file=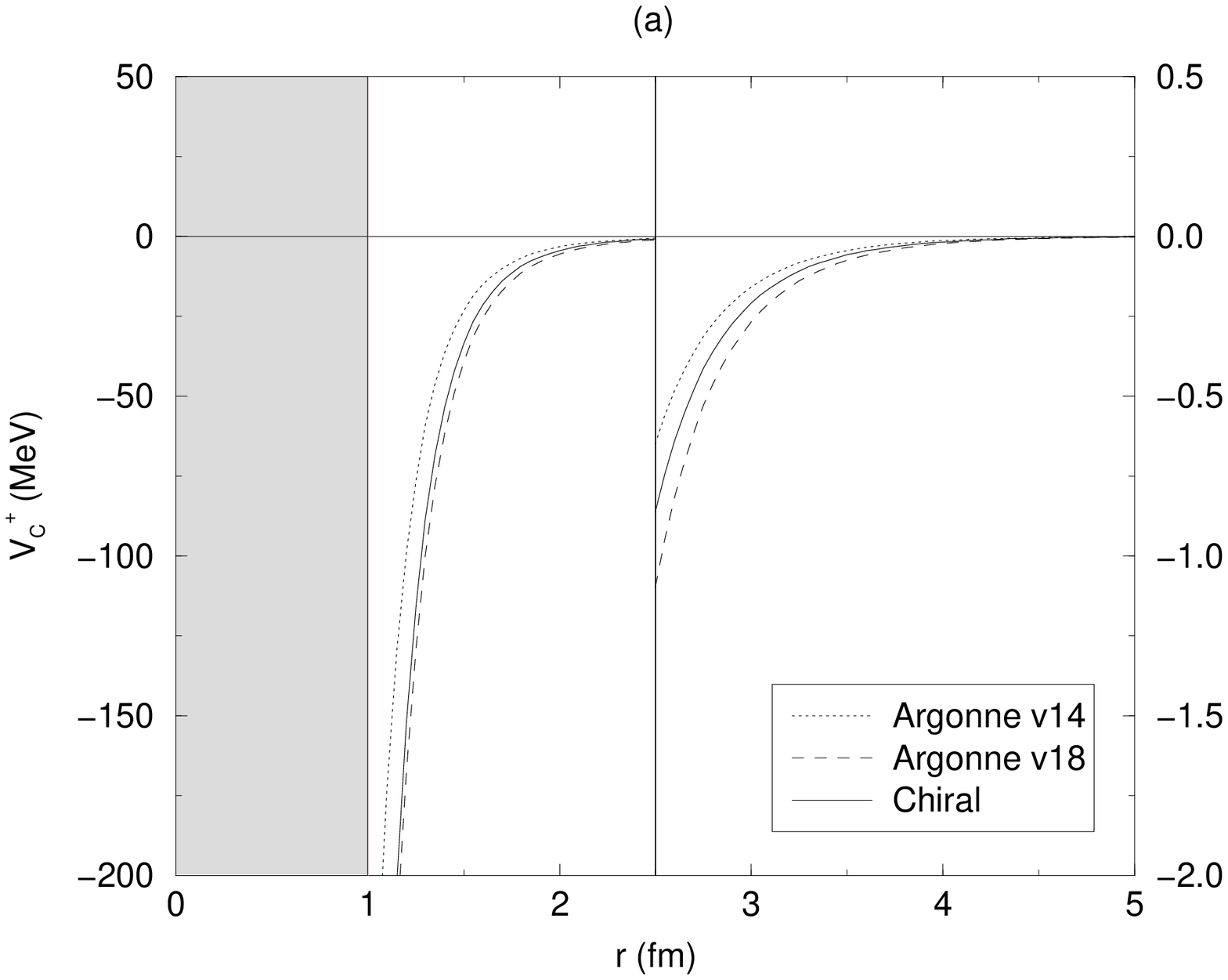,width=190pt}
\epsfig{file=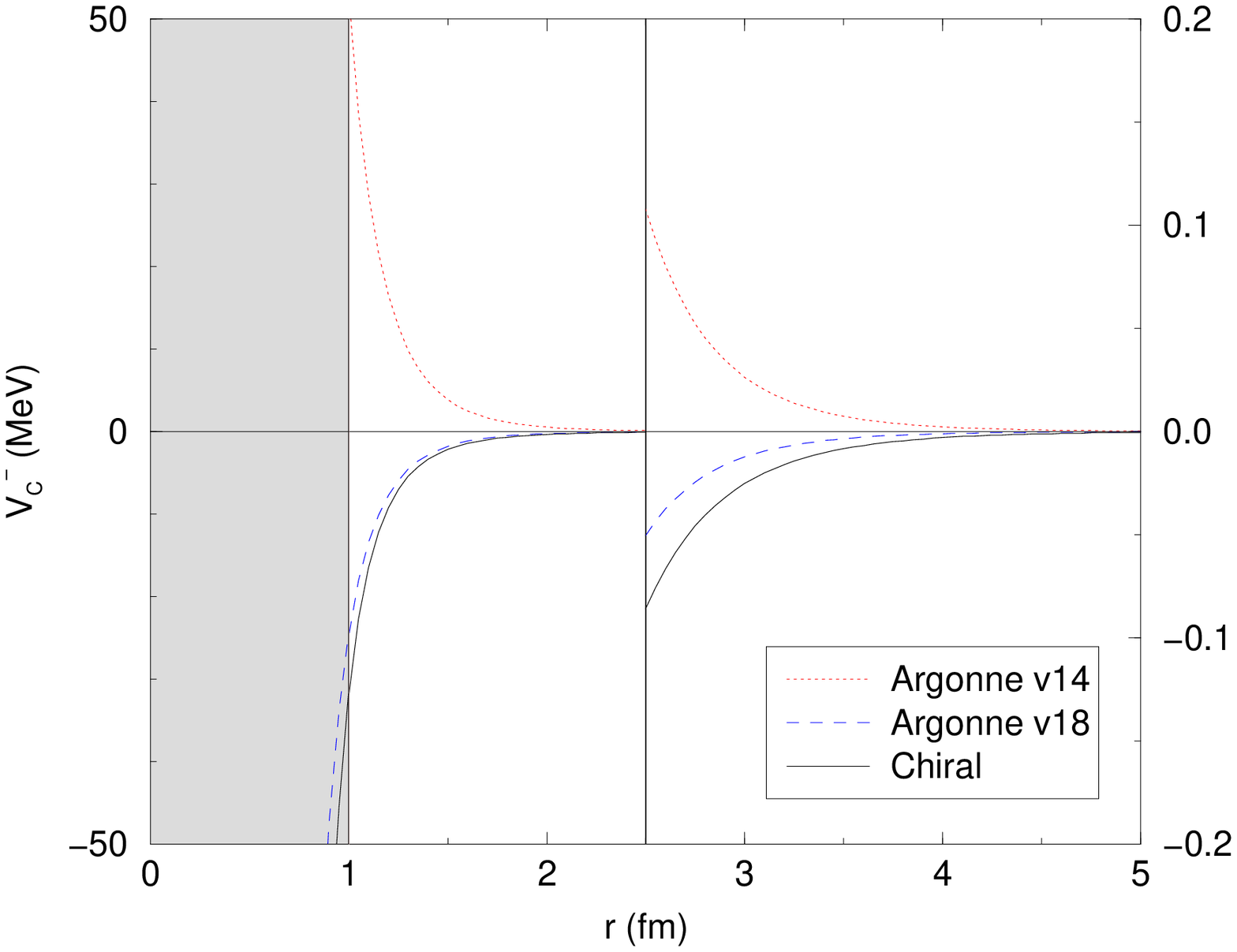,width=190pt}
\caption{Isospin independent $(V_C^+)$ and dependent $(V_C^-)$ 
central components of the $TPEP$\cite{HRR}.}
\end{figure}

\section{Scalar Form Factor}

The discussion in this section is based on results derived in 
Ref.\cite{sigma}.
The nucleon scalar form factor is defined as 
$\la N(p') | \sm \cL_{sb}\, | N(p) \ra = \s_N(t) \; \ub(p')\; u(p) \;,$
where $\cL_{sb}=- \hat{m}\,(\bar{u}u \sp \bar{d}d)$ is the symmetry breaking 
lagrangian. 
The function $\s_N(t)$ can be well represented by a leading $\cO(q^2)$
tree contribution associated with the LEC $c_1$, 
supplemented by two triangle diagrams, involving nucleon and delta
intermediate states, which give rise respectively to $\cO(q^3)$ and
$\cO(q^4)$ corrections. In configuration space contact and loop
contributions split apart and one has
$\tilde{\s}_N(\br) = - 4\, c_1\, \m^2\, \delta^3(\br) + 
\tilde{\s}_{N_N}(r) + \tilde{\s}_{N_\D}(r) \;.$

\begin{figure}[hbt]
\epsfig{file=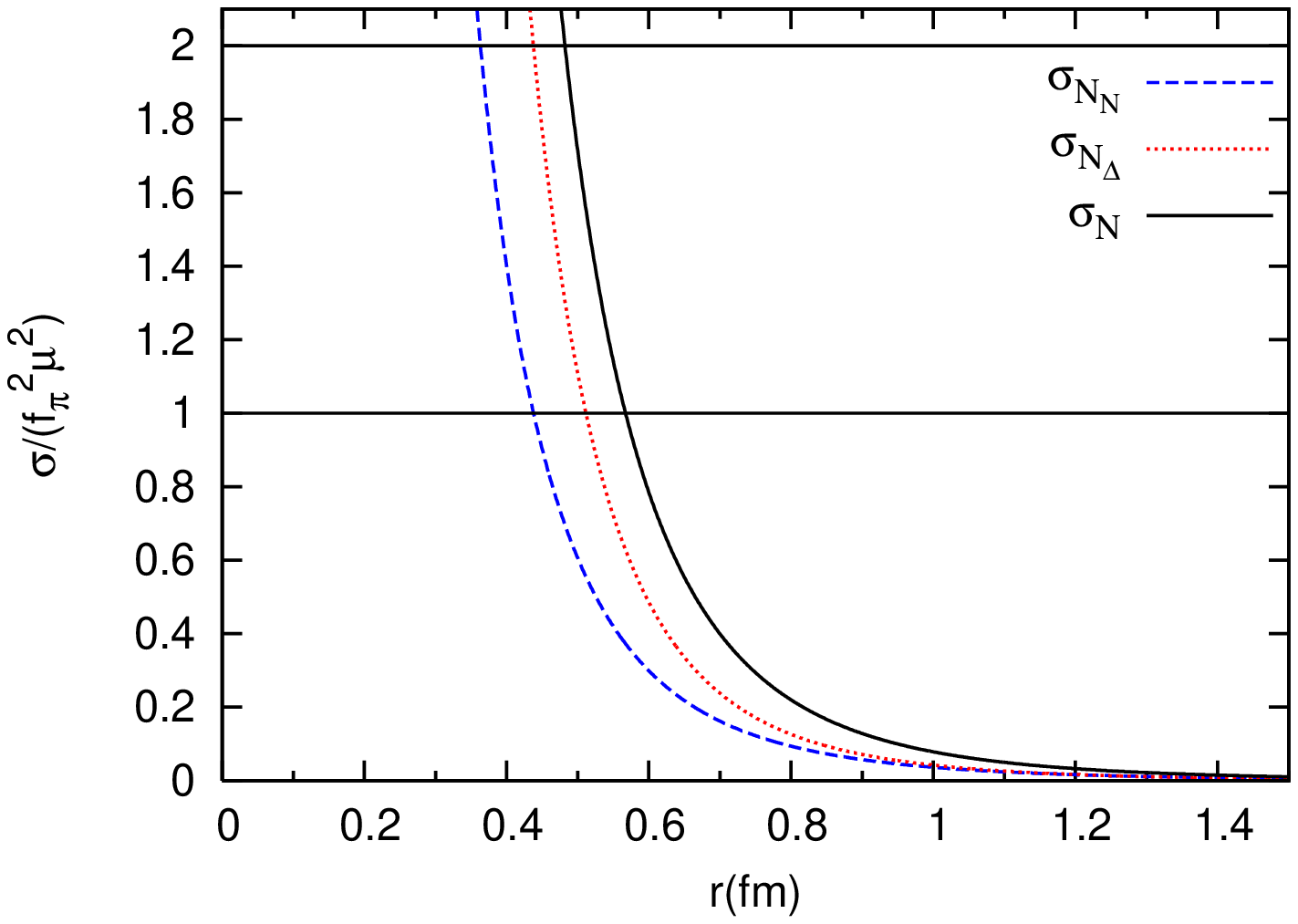,width=135pt,angle=-90}
\hspace{-0.2cm}
\epsfig{file=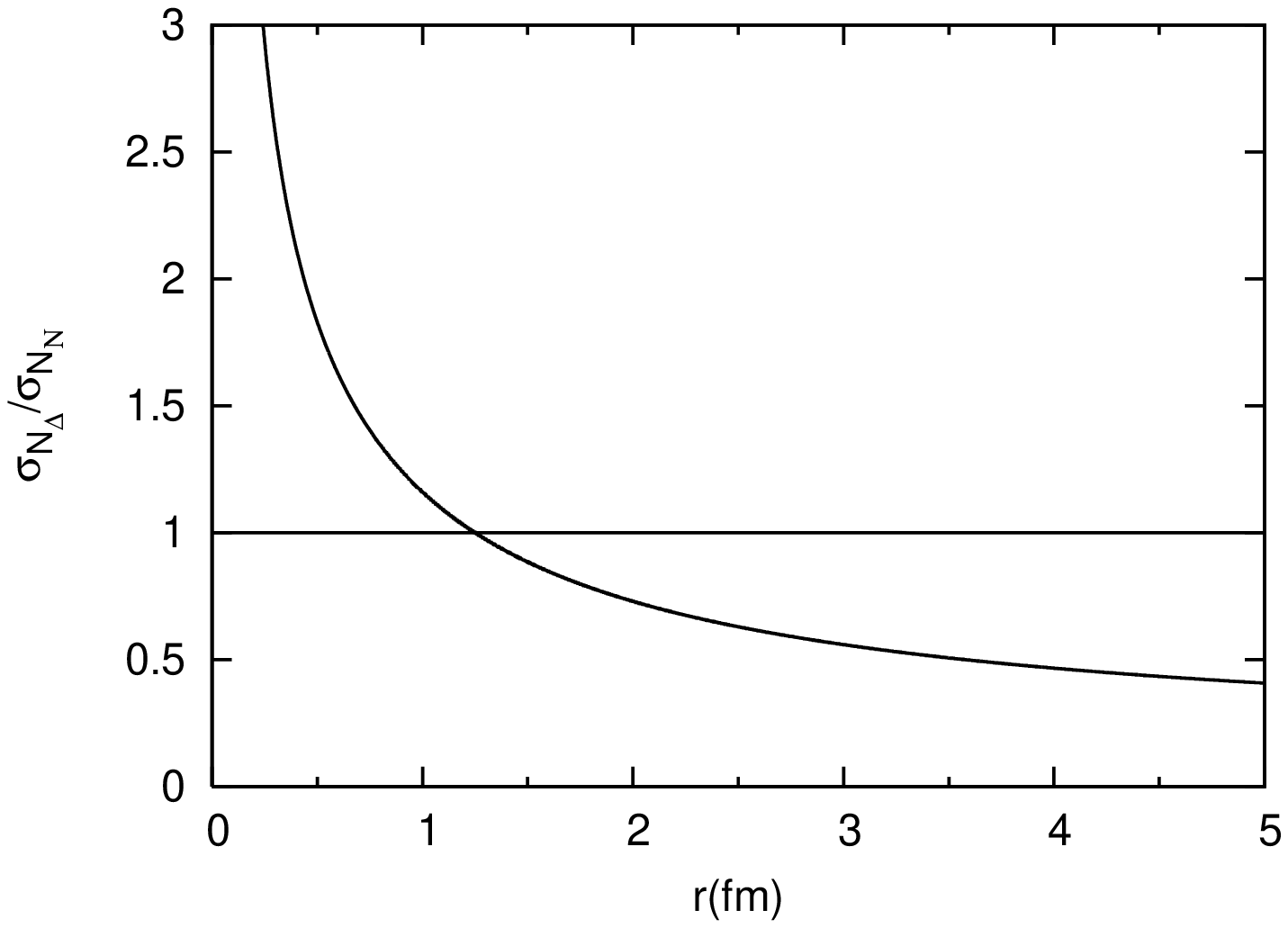,width=135pt, angle=-90}
\caption[]{Ratios $\tilde{\s}_N(r)/(\m^2 f_\p^2)=(1-\cos\theta)$ (left)
and $\tilde{\s}_{N_\D}(r)/\tilde{\s}_{N_N}(r)$ (right).}
\end{figure}

The symmetry breaking lagrangian is written in terms of the chiral angle 
$\theta$ as $\cL_{sb} = f_\p^2 \, \m^2 \,(\cos\theta-1)$, 
in a convention in which the energy density vanishes when $r\rar
\infty$ ($\m$ is the
pion mass).
The ratio $\tilde{\s}_N(\br)/(\m^2 f_\p^2)=(1-\cos\theta)$ is displayed in 
Fig.2(left) and increases monotonically as one approaches the center of the nucleon.
On the other hand, the physical interpretation of the quark condensate 
corresponds to the condition $\la 0|q\bar{q}| 0\ra>0$ and the function 
$\tilde{\s}_N(\br)$ becomes meaningless beyond a critical radius $R$,
corresponding to $\theta = \p/2$. In Ref.\cite{sigma} we assumed that
the condensate no longer exists in the region 
$r<R$ and evaluated the $\p N$ sigma-term using the expression
\beq
\s_N = \frac{4}{3} \p R^3 \;f_\p^2 \m^2
+ 4\p \int_{R}^\infty dr\;r^2\;\tilde{\s}_N(\br)\;.
\label{e2}
\eeq
Depending on the value adopted for the $\p N\Delta$ coupling constant, 
this procedure yielded the range\cite{sigma} 43 MeV$<\s_N<$49 MeV, 
which is fully compatible with the empirical value 45$\pm 8$ MeV.
This indicates that our picture of the nucleon scalar form factor is sound 
and can be used to gain insight about $V_C^+$.

\section{Conclusions}

In the case of the scalar from factor, contributions from $N$ and $\D$
intermediate states are respectively $\cO(q^3)$ and $\cO(q^4)$.
Inspecting Fig.2, one learns the hierarchy predicted by ChPT is subverted
for distances smaller than 1.5 fm, where the delta becomes more important 
than the nucleon.
On the other hand, the good results obtained from eq.(\ref{e2})
for the nucleon sigma-term (and also for $\s_\D$, \cite{sigma}) 
indicate that the functions $\tilde{\s}_{N_N}(r)$ and $\tilde{\s}_{N_\D}(r)$
convey the correct physics at least up to the critical radius $R\sim$ 0.6 fm,
which lies well beyond the domain of the validity of ChPT. 
Just outside this radius the chiral angle is of the order of $\p/2$,
a clear indication of the non-perturbative character of the problem.

This finding is supported by the results from the study of 
the $TPEP$ performed in Ref.\cite{HRR}.
In that work the predictions for $V_C^+$, reproduced in our Fig.1(left), 
were in rather close agreement with the Argonne phenomenological 
potentials\cite{A} up to 1 fm, in spite of the fact that power counting
was ineffective already at 2 fm, as shown in their Fig.6.

Our main conclusion, which needs to be further tested, is that the 
range of validity of calculations based on nucleon and delta intermediate
states is much wider than that predicted by ChPT.

\appendix


\begin{thebibliography}{9}

\bibitem{Bira} Ord\'o\~nez, C. and van Kolck, U.: 
Phys. Lett. B {\bf 291}, 459 (1992).

\bibitem{HB} Entem, D.R. and Machleidt, R.: 
Phys. Rev. C {\bf 66}, 014002 (2002).

\bibitem{RHR} Robilotta, M.R. and da Rocha, C.A.:
Nucl. Phys. A {\bf 615}, 391 (1997); 
Higa, R. and Robilotta, M.R.: 
Phys. Rev. C {\bf 68}, 024004 (2003).

\bibitem{NNN} Ishikawa, I. and Robilotta, M.R.: 
Phys. Rev. {\bf C 76}, 014006 (2007);
Epelbaum, E., contribution to this conference.

\bibitem{HRR} Higa, R., Robilotta, M.R. and da Rocha, C.A.: 
Phys. Rev. C {\bf 69}, 034009 (2004).

\bibitem{sigma} Cavalcante, I.P., Robilotta, M.R., S\'a Borges, J., 
Santos, D.O., and Zarnauskas, G.R.S., 
Phys. Rev. C {\bf 72}, 065207 (2005).

\bibitem{A} Wiringa, R.B., Smith, R.A., and Ainsworth, T.L.:
Phys. Rev. C {\bf 29}, 1207 (1984);
Wiringa, R.B., Stocks, V.G.J., and Schiavilla, R.:
Phys. Rev. C {\bf 51}, 38 (1995).


\end{thebibliography}
\end{document}